# Unveiling the Thermal and Aqueous Stability of 1D Lepidocrocite Titania

*Risha A. Iythichanda [a,b], Sukanya Maity [a], Mustafa M. Aboulsaad [c], Tomas Edvinsson [c], Johanna Rosen [a,b], and Per O.Å. Persson * [a,b]*

[a] Department of Physics, Chemistry and Biology (IFM), Linköping University, 58183 Linköping, Sweden

[b] Wallenberg Initiative Materials Science for Sustainability (WISE), Linköping University, Department of Physics, Chemistry and Biology (IFM), 58183 Linköping, Sweden

[c] Department of Materials Science and Engineering; Solid State Physics, Uppsala University, 75105 Uppsala, Sweden

* Corresponding Author

E-mail address: per.persson@liu.se (Per O.Å. Persson)

**ABSTRACT:** One-dimensional (1D) lepidocrocite titanium dioxide (TiO$_2$) filaments are investigated with respect to their thermal and aqueous stability. Structural and phase evolution are examined using *in situ* heating in vacuum within (scanning) transmission electron microscopy combined with electron energy loss spectroscopy, and at ambient conditions using Raman spectroscopy. The filaments retain their lepidocrocite structure up to ~300 °C, above which localized sintering and amorphization occur at filament overlap junctions. With further heating, the amorphous regions crystallize into anatase TiO$_2$, with Raman spectroscopy corroborating the onset of structural disorder. Long-term aqueous storage (>100 days) at ambient conditions induces transformation into flake-like anatase nanoparticles. This process is strongly suppressed under refrigerated storage, where no structural changes are observed over the same period. These results establish critical thermal and environmental stability thresholds that define operational advantages and limits for emerging applications of 1D lepidocrocite TiO$_2$ filaments.

Over the past decades, low-dimensional materials (LDMs) such as two-dimensional (2D) [1] and, more recently, one-dimensional (1D) materials [2] have attracted significant attention in material science. The reduced dimensionalities give rise to pronounced changes in fundamental properties like quantum confinement effect,[3] structural anisotropy,[4] photon transportation,[5] mechanical flexibility [6] and offers a new platform for tuning electronic,[7] optical [8] and catalytic properties.[9] 1D nanostructures exhibits a unique opportunity as a consequence of the further enhanced aspect ratio,[10] leading to staggering gravimetric and volumetric surface area,[11] which renders them highly attractive for applications in, e.g. energy storage,[12] catalysis,[9, 13] optoelectronics,[14] water purification,[15] and biomedical systems.[16] These opportunities make LDMs prime candidates for next generation functional materials targeting the UN global goals.[17]

1D lepidocrocite structure titanium dioxide (filaments),[18] stand out among the recently discovered LDMs due to their exceptional structural properties. The dimensions of these filaments comprise lengths beyond 1 µm, a width of typically 3-6 nm, and a thickness of a single unit cell. These ribbons only grow in a single direction and are crystalline throughout the filament. The filaments bend significantly, and exhibit a high amount of single atom vacancies, such as Ti- and O-vacancies. Furthermore, the edges of the filaments exhibit a disordered structure.[19] Their geometry thus results in an exceptional surface to volume ratio, a cotton-like material with high permeability, a large bandgap, and tunable electronic characteristics. While $TiO_2$ in its anatase and rutile forms has been extensively studied for a range of applications,[20] much less is known about these filaments in terms of fundamental properties and applications they are still largely unexplored. However, the filaments present a unique combination of properties that are essential for the development of more efficient, sustainable and scalable technologies.[18]

Previous reports have largely focused on the synthesis, morphology, and immediate properties of these filaments, with thus far limited attention given to their stability, despite their novelty and prospects. While $TiO_2$ in other forms has been widely studied in terms of stability and phase transitions,[21,22] there remains a gap in the understanding of stability of filaments under both thermal and environmental conditions. This is of crucial practical concern, as their performance in real world applications depends on their ability to maintain structural integrity over time.

This report presents a systematic investigation of the thermal stability and phase evolution of these filaments in vacuum, focusing on their behaviour under controlled heating (up to 600 °C), as well as their stability during prolonged storage under ambient and refrigerated conditions. By performing *in situ* heating, inside the vacuum of a (scanning) transmission electron microscopy (TEM) combined with electron energy loss spectroscopy (EELS), and during Raman spectroscopy under ambient conditions, we provide comprehensive insights into the thermal stability. Beyond that we also explore the stability of the filaments during extended storage in aqueous conditions. Accordingly, this study fills an important gap in the understanding of the filaments and provides boundary conditions for their practical application throughout various fields.

**RESULTS**

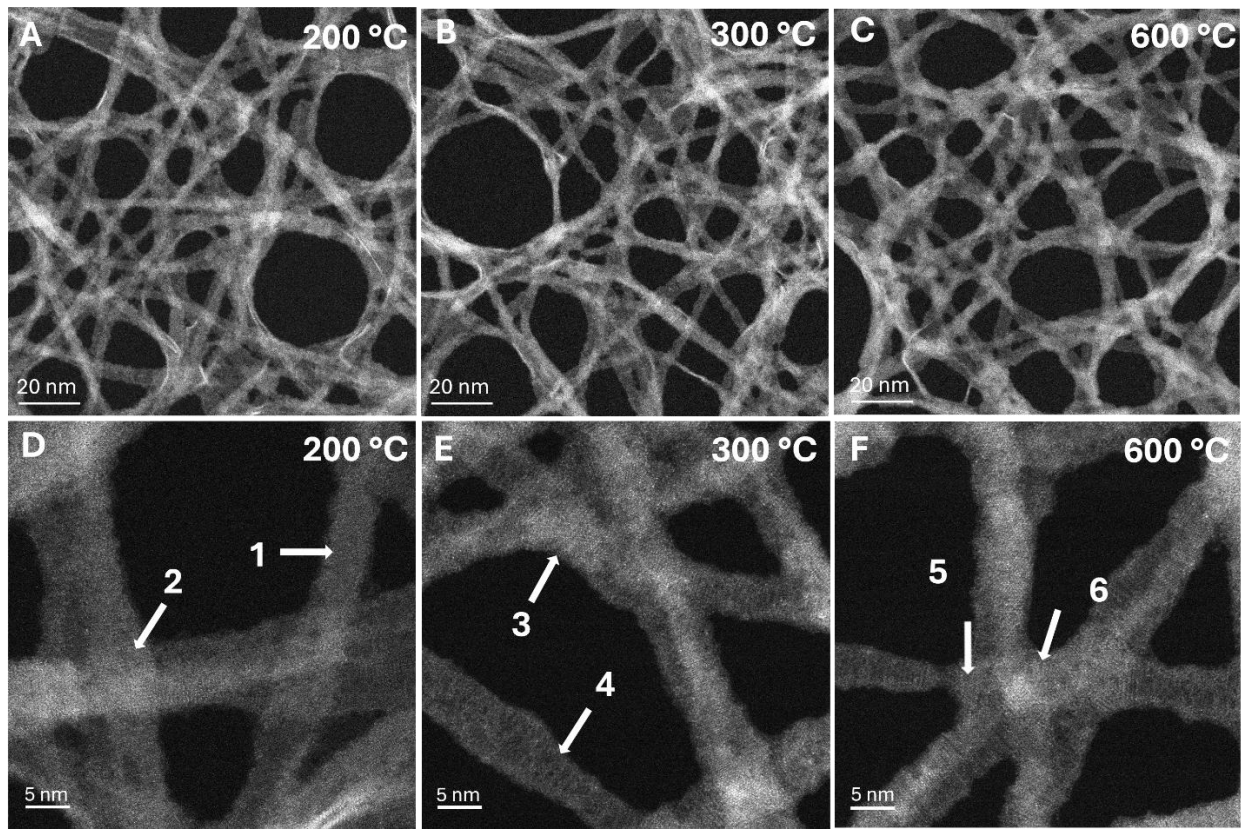

**Figure 1.** Overview and high magnification STEM images of the *in situ* thermal stability experiment at 200 °C (A, D), 300 °C (C, F), and 600 °C (B, E), respectively.

The thermal stability of the filaments in vacuum was investigated using *in situ* heating in a (S)TEM. 1D lepidocrocite $TiO_2$ filaments were stored in aqueous solution and drop casted on a MEMS heating chip. Accordingly, deposited filaments are randomly organized and overlap between two or more filaments occurs spontaneously. This can be seen in the STEM images in Figure 1, where the mass-thickness dependent contrast results in brighter regions where filaments overlap, compared to regions with single filaments or vacuum (dark). To avoid contamination by species from the microscope vacuum, the samples were first heated to, and imaged at 100 °C, at which point the filaments exhibits their pristine morphology, see Figure S1. Upon heating to 200 °C, the filaments maintain their structural integrity as demonstrated in Figure 1A,D. The overview image shows an intricate network of filaments, while the high magnification image shows single filaments with homogeneous contrast (constant thickness and mass), see label 1, Figure 1D. In regions where filaments overlap, the boundaries can be clearly seen and contrast varies with integer steps (dependent on number of filaments projected in the direction of the e-beam), see label 2 in Figure 1D.

Upon heating to 300 °C, the overview image in Figure 1B shows no distinct changes. However, upon closer inspection of the high magnification image in Figure 1E initial changes in the overlapping regions become apparent. The close contact between two or more filaments results in a localized increase in thickness, accompanied by structural disorder at the edges. Finally, it is no longer possible to distinguish one filament from another, indicating signs of local sintering (label 3, Figure 1E). In overlapping regions, the sintering results in localized crystalline disorder. In contrast, the single filaments remain largely unaffected, and the lepidocrocite phase is preserved (label 4, Figure 1E).

With further increase of annealing temperature to 400 °C and 500 °C (Figure S1(B, D) and (C, E) respectively), a further pronounced structural transformation is evident. The lepidocrocite phase persists in the non-overlapping regions (label 1, Figure S1D), while sintering and amorphization continue in regions where filaments overlap (label 2, Figure S1D). Notably, at 500 °C the amorphous regions begin to re-crystallize (label 4, Figure S1E), whereas the single filaments (label 3, Figure S1E) display local atomic defects (vacancies) that disrupts the regular atomic arrangement in the filaments and can also be seen to form local vacancy clusters.

After heating to 600 °C significant morphological and structural changes occur to the point that even the overview image exhibits a duller appearance compared to the overview images at lower temperatures. From the high magnification image in Figure 1F, it can be observed that the overlapping regions of the filaments show pronounced recrystallization, where the amorphous phase has transformed into anatase (label 6, Figure 1F). Simultaneously, the single filaments begin to degrade. This can be seen e.g. by necking of the filaments (label 5, Figure 1F), and by an enhanced contrast near the filament edges which is indicative of amorphous material that exhibits a higher thickness compared to the otherwise unit cell thick filament. Together these are indications of thermal breakdown with a projected conversion from the lepidocrocite phase into anatase $TiO_2$. The filaments were investigated in parallel by electron energy loss spectroscopy (EELS) throughout the temperature range, providing insights into changes in the local chemistry Spectra were acquired by averaging information across a filament rich region, as extend exposure to the high energy electrons in localized regions otherwise affects the structure of the filaments. Figure 2 follows the Ti $L_{3,2}$ and O K-edges throughout the heating process. The spectra remain largely unchanged from 200 to 500 °C, showing no significant variation in fine structure, peak positions, or relative intensities. This spectral stability is consistent with the structural observations made by STEM. The lack of any distinct spectral shifts reinforces the conclusion that the lepidocrocite phase remains stable up to 500 °C in regions without overlap. However, at 600 °C, both the Ti $L_{3,2}$ and the O K-edges exhibit notable differences with respect to the lower temperatures. For the Ti $L_{3,2}$ edge, this is most prominently visible by the emerging separation of the $t_{2g}$ and $e_g$ crystal field splitting features within both the $L_2$ and $L_3$ edges, accompanied by a small but noticeable ~0.25 eV chemical shift towards higher energy of the edge features. The edge thus assumes characteristic profile of anatase $TiO_2$, which enhances the interpretation of crystallization of the amorphous phase. Also, the O K-edge at 600 °C reveals the appearance of a distinct hybridization feature, and a more pronounced chemical shift, associated with Ti 3d and O 2p orbital interactions which are also typical for the anatase structure.

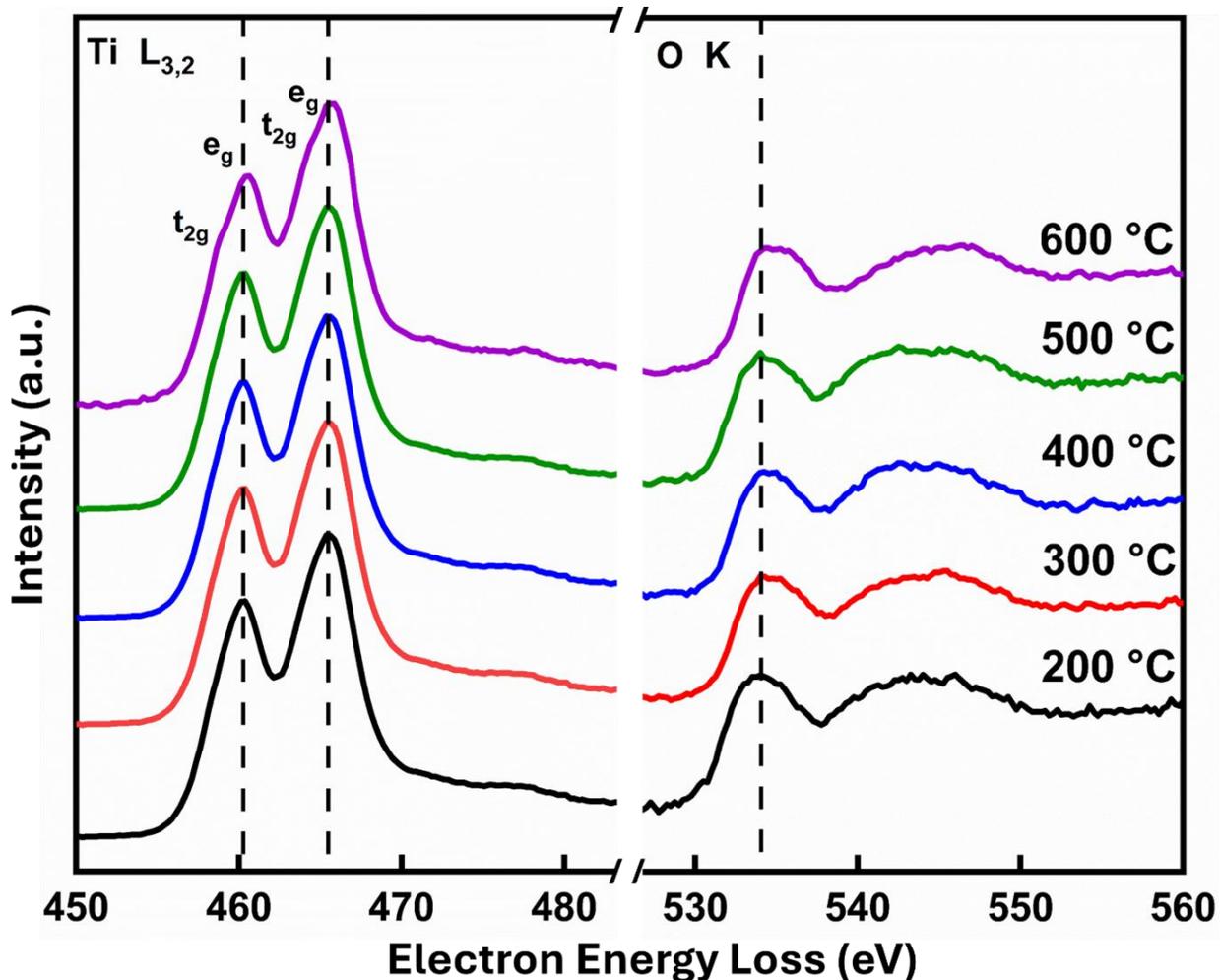

**Figure 2:** EELS core-loss spectra of the Ti $L_{3,2}$ and O K-edges of the filaments after heating to temperatures ranging from 200 to 600 °C

To exclude electron beam induced artefacts and confirm that the observed changes arise from thermal effects, we performed *in situ* Raman spectroscopy from room temperature (RT) up to 300 °C see

Figure *3*, following the same heating protocol used during the *in situ* STEM experiment. Unlike STEM, the Raman measurements were conducted under ambient atmospheric conditions

rather than in vacuum. At RT, eleven characteristic Raman peaks are observed at 115, 189, 275, 288, 388, 449, 699, 755, 841, 920, and 951 cm⁻¹, consistent with layered lepidocrocite titanates.[22] The bands in the 600–940 cm⁻¹ region are assigned to Ti–O vibrations.[23] In particular, three Ag modes at 269, 434, and 824 cm⁻¹ indicate a well-developed layered lepidocrocite framework.[24] The absence of the $E_g$ mode at 144 cm⁻¹, characteristic of anatase $TiO_2$, further corroborates the

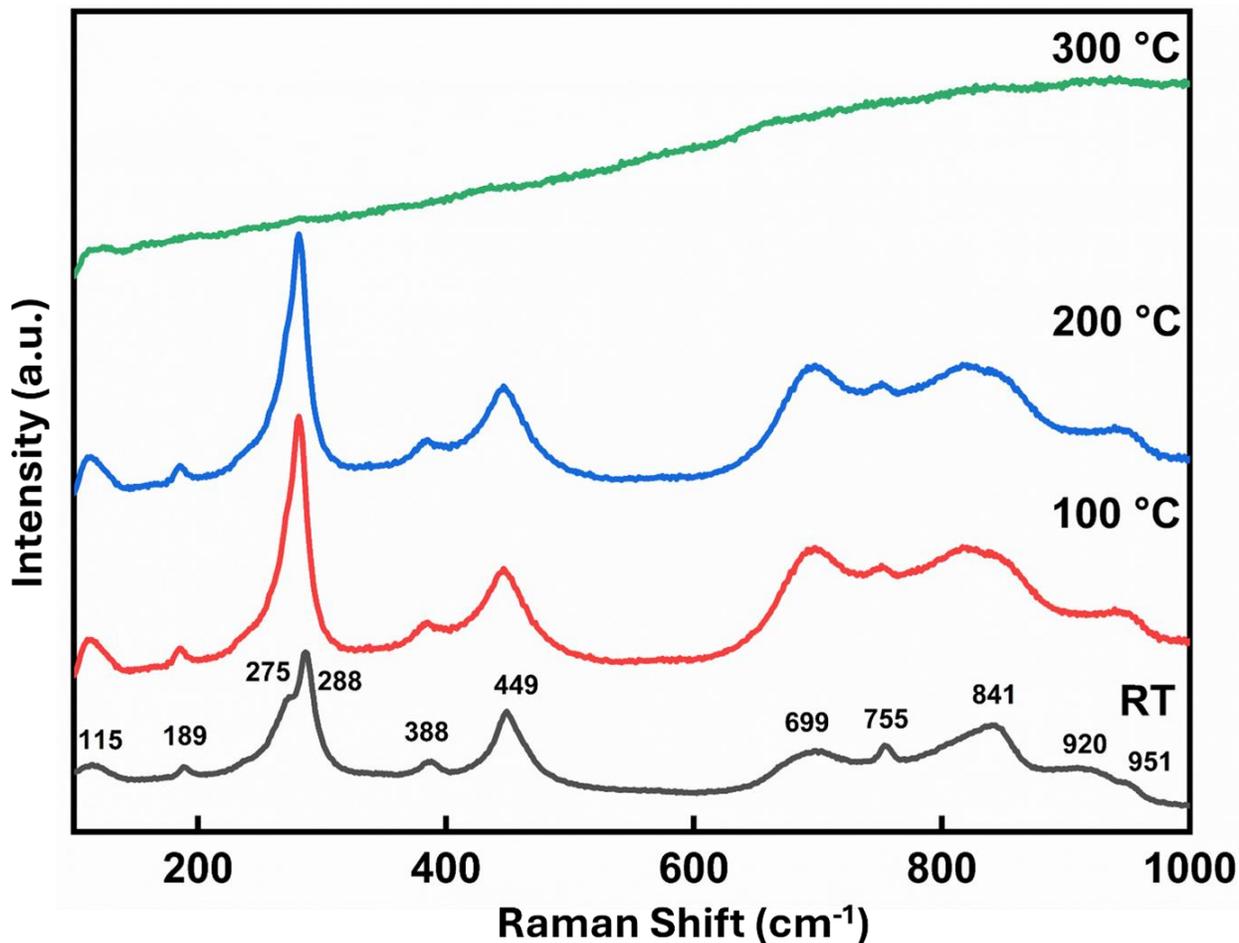

lepidocrocite phase purity.[25]

**Figure 3:** *In situ* Raman spectra of 1D Lepidocrocite Titania at different temperatures.

Moreover, the lack of a 665 cm⁻¹ Raman band confirms the absence of the 2D Lepidocrocite $TiO_2$

phase.[26] As this feature originates from the ~ 700 cm⁻¹ Ag band when 2D Lepidocrocite TiO₂ co-exists with 1D Lepidocrocite TiO₂. With increasing temperature, the spectra remain characteristic of lepidocrocite TiO₂ and show no significant evolution. The diagnostic peaks of orthorhombic lepidocrocite at 187, 280, 382, 446, 799, and 951 cm⁻¹ persist throughout the entire temperature range.[24,27] At 300 °C, however, the spectra are dominated by a broad fluorescent background. This is consistent with our prior *in situ* heating STEM observations, wherein overlapping regions began to sinter at 300 °C. For the *in situ* Raman measurements, the specimen was prepared by drop-casting multiple filament layers onto a glass substrate, producing extensive overlap. Upon heating to 300 °C, these overlapping regions interact and generate amorphous regions across the network, which in turn yield the observed fluorescence. Collectively, these Raman data support the conclusion that the structural responses are thermally driven rather than beam induced.

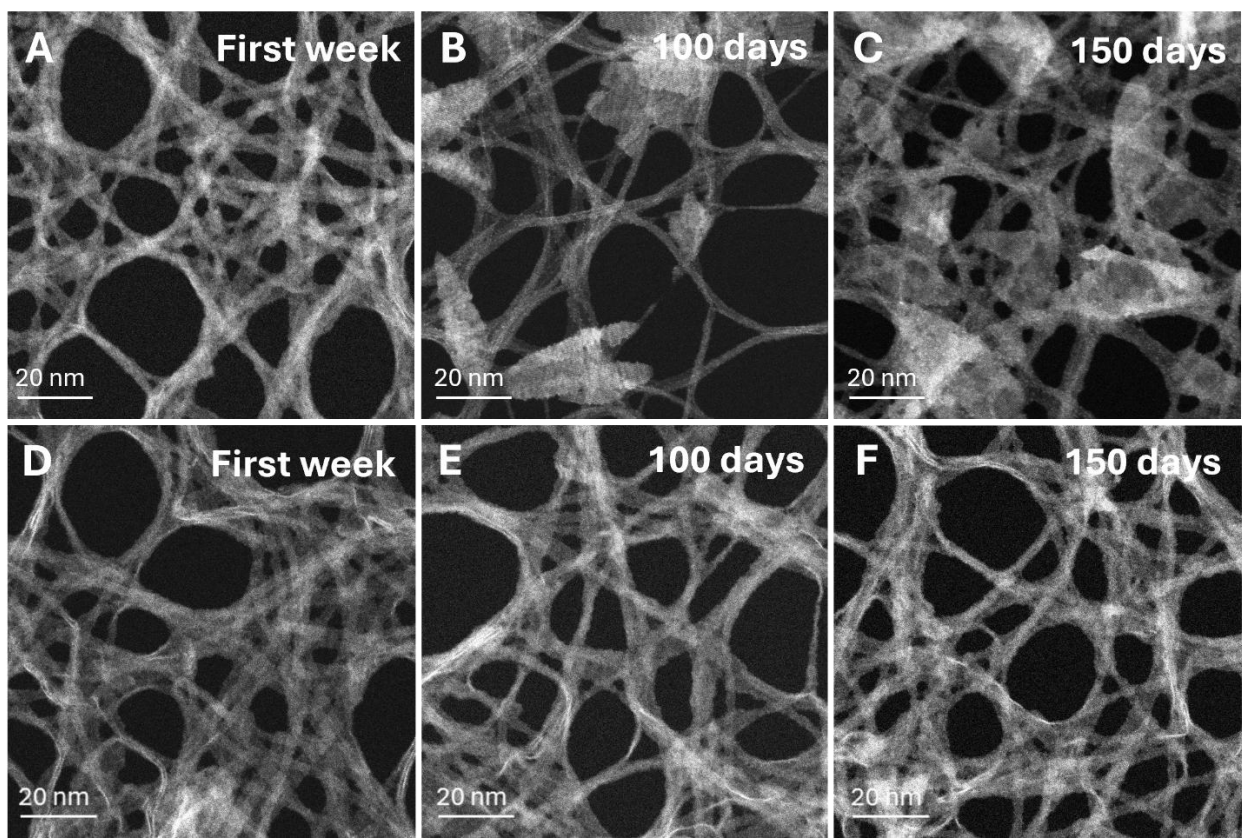

**Figure 4.** Aqueous Stability of 1D Lepidocrocite Titania stored at ambient (A, B, C) and refrigerated conditions (D, E, F), imaged at 200 °C.

Following the thermal stability experiments, the filaments were dispersed in a colloidal solution (4mg/ml) and stored under ambient and refrigerated conditions in the dark to monitor their long-term stability. As shown in Figure 4A, the filaments exhibited structural stability during the first week. However, with increasing storage duration, the gradual formation of flakes was observed. As shown in Figure 4B, showing the filaments after 100 days of storage, the presence of flake-like particles becomes increasingly evident, and after 150 days, shown in Figure 4C, the colloidal solution was densely populated with flakes, as further illustrated in the overview images Figure S2A, B. The flakes were identified as single crystal $TiO_2$ anatase particles, as demonstrated by high resolution scanning transmission electron microscopy (HRSTEM) analysis see Figure S3.

Also for these samples, EELS was employed. At the early stages of the experiments, Ti $L_{3,2}$ edge spectra are characteristic of the lepidocrocite phase see Figure S4. However, by 100 days, significant changes emerge in the spectral features, particularly the appearance of the distinct $t_{2g}$ peaks, which are clear signatures of anatase formation. Additional confirmation is provided by the oxygen K edge see Figure S4. Around the 100 days mark, distinct hybridization features develop, reflecting a transformation in the local oxygen coordination environment to anatase $TiO_2$ and signify the progressive conversion of lepidocrocite filaments over time.

In parallel, an identical colloidal solution (4mg/ml) was stored under refrigerated conditions (4 °C) and monitored over the same period Figure 4(D, E, F). In this case no formation of competitive phases nor degradation of the filaments was observed throughout the period, indicating that the reduced temperature significantly suppresses the transformation process. This finding clearly underscores that increased temperature or environmental conditions facilitates the degradation and phase transition.

Notably, the high synthesis yield ($\sim$98-99%) ensures that nearly all precursor material is converted into lepidocrocite and that there are negligible unreacted material or byproducts available to account for the formation of the anatase particles. Accordingly, the appearance of flakes can be inferred to occur as the result of degradation of the filaments under ambient conditions, due to an intrinsic instability and transformation of the filaments over time. Interestingly, despite the internal transformation, no visible changes in the physical appearance (e.g., colour, turbidity or precipitation) of the colloidal solution were observed during the 150-day period Figure S5A,B emphasizing the need for microscopic and/ or spectroscopic techniques to detect the changes.

## DISCUSSION

The thermal evolution of lepidocrocite $TiO_2$ filaments under *in situ* heating reveals a complex interplay of structural stability, defect formation, and phase transformation, which depends on filaments morphology and arrangement. Notably, the sintering observed in the overlapping regions is presumably a consequence of the systems tendency to minimize its total surface and interfacial energy at elevated temperatures. When two or more filaments are in close contact, the localized increase in thickness, combined with the observed disorder at the edges of the filaments, promotes atomic diffusion, enabled by the applied thermal energy. In contrast, the non-overlapping filaments demonstrate stability up to $\sim$500 °C. The absence of contact between filaments removes the driving force for sintering. However, these regions are not unaffected as structural defects begin to emerge in the filament core as the temperature increases. These defects, most likely involving titanium and oxygen vacancies, are presumably thermally activated point defects, likely to form at elevated temperature inside the (S)TEM vacuum. The formation of vacancies and vacancy clusters weakens the crystal lattice and is a well-documented behaviour in $TiO_2$.[28,29] As vacancies accumulate, they increase the probability for lattice disruption and leads to increasing crystalline disorder that eventually results in structural breakdown, even in the absence of overlap between filaments.

The most significant structural transformation occurs at 600 °C, where sintered amorphous regions of the filaments undergo recrystallization into the anatase phase. This transition aligns with known behaviours of titania materials under thermal treatment, while the formation is not simply due to thermal melting, but a thermally driven phase transition.[30] During this process, the atomic mobility increases, especially at higher temperatures, allowing atoms to shift into more stable phase.[31] In non-overlapping regions at 600 °C, the emerging defects increase the crystalline disorder that eventually leads to structural breakdown, with fractures and lattice misalignments becoming more apparent.

When *in situ* Raman spectroscopy was performed on the filaments, the Raman activity was hidden by competing photoluminescence signal as early as 300 °C in contrast to TEM (which shows complete structural degradation at 600 °C). This can be rationalized by differences in sensitivity of the technique but a part also to the sample preparation and resulting filament density in the samples. In the *in situ* Raman experiment, the sample was prepared by drop casting multiple layers of the filaments, resulting in a denser film, consequently exhibiting more overlap compared to the sample prepared for TEM. The broad fluorescent background in the Raman spectra at 300 °C is indicative of the start of the structural breakdown occurring at the same temperature as for the initial observations of sintering in the TEM. Therefore, the results are not contradictory but rather reflect the sensitivity of the technique to bulk and surface defect formation in a denser and more thermally responsive configuration. The difference in atmosphere (ambient air in Raman vs.

vacuum in TEM), spatial averaging (Raman probes a micrometre scale area vs. an atomic scale in TEM), and sample architecture collectively explain the spectral changes observed in Raman.

The revealed thermal behaviour of lepidocrocite $TiO_2$ filaments provides key insights into the potential applications and limitations. A thermal stability up to 300 °C suggests suitability for applications such as photocatalysis, low temperature sensors, energy storage components, and transparent coating in optoelectronics. Maintaining phase purity structural integrity in this range is crucial for optical, electronic, and catalytic performance. However, the observed phase transition and structural changes at higher temperatures limit their use in higher temperature processes, thermal cycling, or extreme conditions.

Moving to aqueous stability, the lepidocrocite filaments remains structurally intact and colloidally robust for an extend period (~100 days) under ambient conditions. Throughout this timeframe, the material preserves their characteristic 1D morphology and visual appearance, suggesting that the material is sufficiently stable for most practical processing and application workflows. This long stability window is particularly encouraging, only after prolong storage beyond ~100 days the structural transformation to anatase is observed in ambient condition. Notably, this transformation progresses slowly and can be effectively supresses by storing the colloidal solution under refrigerate conditions Overall the filaments stability makes them well suited for application that capitalize on the unique 1D morphology, such as colloidal inks, photocatalysis, biosensing, and electrochemical devices. The material performs reliable within the operational and processing timescale typical for these applications. Although exceedingly long term ambient storage eventually induces partial transformation, simple strategies such as cold storage and surface modification are sufficient to preserve structural integrity. Thus, the observed behaviour does not diminish their practical utility rather, it provides valuable guidance for handling and long-term preservation.

**CONCLUSION**

This study provides insight into the thermal and aqueous stability of 1D lepidocrocite $TiO_2$ filaments using *in situ* STEM, EELS, and *in situ* Raman spectroscopy. A film of filaments remains thermally stable below 300 °C, retaining its lepidocrocite phase and structural integrity. Sintering between overlapping filaments, resulting in amorphization, is observed in STEM from 300 °C onwards. However, individual filaments demonstrate remarkable thermal robustness and remain stable up to >500 °C. A distinct phase transformation from amorphous into anatase $TiO_2$ occurs at ~600 °C, as confirmed by both imaging and spectroscopy. In an aqueous colloidal system, the filaments remain structurally stable for the initial period under ambient condition, but gradually transform into anatase $TiO_2$ over 150 days, as evidenced by the appearance of anatase flakes, which is attributed to the intrinsic instability of the lepidocrocite phase at ambient conditions. Conversely, a sample stored under refrigerated condition showed no signs of degradation or phase change, highlighting the critical role of temperature for long term colloidal stability. In summary, this study reveals the thermal and environmental stability and limitations for the 1D lepidocrocite filaments and offers crucial insights for the application of this material.

## EXPERIMENTAL METHODS

Sample preparation

Filaments were prepared according to previously published methods.[18]

Materials: Titanium (IV) oxysulfate, (TiOSO$_4$) ($\geq$ 29% wt.% Ti, anhydrous basis, powder) and Tetramethylammonium hydroxide, (TMAH) (25 wt.%), were procured commercially (Sigma Aldrich). Ethanol, (EtOH) 200 proofs, is obtained from VWR Chemicals.

Experimental Details:

The one-pot synthesis method for creating 1D lepidocrocite TiO$_2$ filaments involves mixing the TiOSO$_4$ powder in the TMAH solution in Ti:TMA molar ratio to 0.4 at temperatures of 80 °C. Firstly, TiOSO$_4$, in powder form, was added to the TMAH solution in polyethylene bottles into which a hole was poked to prevent any pressure buildup. The system was kept on a magnetic stirrer in an oil bath for 2 days. When the TiOSO$_4$ dissolved in the TMAH a clear solution was obtained. After the reaction, a white sediment was obtained that was in turn washed with EtOH several times until the pH was $\approx$ 7 to remove the excess TMA and then centrifuged until the supernatant was clear. After discarding the supernatant, the precipitate was dispersed in DI water and shaken until a clear, transparent colloid was formed. This colloid solution is further used for characterizations.

Sample preparation for in situ STEM and *in situ* Raman spectroscopy:

The filaments for the *in situ* STEM investigation were drop casted from an aqueous dispersion on SiN$_x$ MEMS heating chips, The MEMS chip was mounted on the DENSsolutions holder and inserted into the TEM, where the sample was heated at 100 °C overnight in a vacuum environment to eliminate any potential contamination. The thermal stability of these filaments was studied up to 600 °C, with the temperature interval of 100 °C and a duration of 15 minutes at each temperature step. After each heating step, the sample was returned to 100 °C for imaging and EELS analysis. while for Raman spectroscopy, a thin film of the filaments was prepared by (drop casting ~5mg of material on a glass substrate). The long-term storage experiments were conducted by placing colloidal solutions (4mg/ml) of filaments under ambient and refrigerated conditions. In the experiment the initial procedure is same as thermal stability test but here all the imaging and EELS analysis was carried out at 200 °C.

Sample Characterization:

Scanning TEM (STEM) high-angle annular dark-field (STEM-HAADF) and EELS were performed using a double-corrected FEI Titan$^3$ 60–300, operated at 300 kV. For the *in situ* thermal characterization, a MEMS based heating chip (DENSsolutions) was mounted in a DENSsolutions Wildfire heating holder. The chip consists of an electron transparent SiN membrane with an integrated resistive micro heater. To enhance imaging quality, a focused ion beam (FIB) system (Thermo Scientific Helios 5 UC DualBeam) was employed to prepare holes in the SiN windows, improving electron transparency and resolution. The heating holder ensured stable electronical contact with the MEMS device and unable the observation of structural and morphological evolution at different temperatures, while minimizing thermal drift and maintaining high spatial resolution.

Raman spectra were acquired using a Renishaw inVia Reflex confocal microscope, equipped with a frequency doubled Nd:YAG laser operating at 532 nm with a maximum power of 43 mW at the sample surface. The single spot and mapping spectra were collected using 50x long working distance objective (NA 0.5), 2400 lines/mm grating, and a laser power adjusted to 10 % of the maximum power (4.3 mW) using neutral density filters to avoid any deterioration from the laser heating. The spectra were recorded with 20 seconds integration time and 5 accumulations, covering a static scan range of 85–1345 cm$^{-1}$. Prior to spectra acquisition of the samples, the spectrometer was calibrated with measurements on silicon, confirming that the characteristic Si peak is found at 520.5 cm$^{-1}$.

## ASSOCIATED CONTENT

Supporting information is available free of charge.

Overview and high magnification STEM images of the *in situ* thermal stability experiment at 100 °C 400 °C  and 500 °C, Aqueous Stability of 1D Lepidocrocite Titania: Overview STEM images, HRSTEM image of anatase TiO$_2$ flake, Aqueous stability of 1D Lepidocrocite Titania EELS core-loss spectra Ti $L_{2,3}$-edge and O $K$-edge, Physical appearance of 1D lepidocrocite Titania in room temperature (PDF)

## AUTHOR INFORMATION

### Corresponding author

Per O.Å. Persson

Department of Physics, Chemistry and Biology (IFM), Linköping University, 58183 Linköping, Sweden

Wallenberg Initiative Materials Science for Sustainability (WISE), Linköping University, Department of Physics, Chemistry and Biology (IFM), 58183 Linköping, Sweden

 per.persson@liu.se.

**Authors**

Risha Achaiah Iythichanda - Department of Physics, Chemistry and Biology (IFM), Linköping University, 58183 Linköping, Sweden

Wallenberg Initiative Materials Science for Sustainability (WISE), Linköping University, Department of Physics, Chemistry and Biology (IFM), 58183 Linköping, Sweden

E-mail: risha.achaiah.iythichanda@liu.se,

Johanna Rosen - Department of Physics, Chemistry and Biology (IFM), Linköping University, 58183 Linköping, Sweden

Wallenberg Initiative Materials Science for Sustainability (WISE), Linköping University, Department of Physics, Chemistry and Biology (IFM), 58183 Linköping, Sweden

E-mail: johanna.rosen@liu.se

Sukanya Maity - Department of Physics, Chemistry and Biology (IFM), Linköping University, 58183 Linköping, Sweden

E-mail: sukanya.maity@liu.se

Mustafa Mahmoud Aboulsaad - Department of Materials Science and Engineering; Solid State Physics, Uppsala University, 75105 Uppsala, Sweden

E-mail: mustafa.aboulsaad@angstrom.uu.se,

Tomas Edvinsson - Department of Materials Science and Engineering; Solid State Physics, Uppsala University, 75105 Uppsala, Sweden

E-mail: tomas.edvinsson@angstrom.uu.se.

**AUTHOR CONTRIBUTIONS**

Risha Achaiah Iythichanda conceived and planned the experiments, performed all *in situ* microscopy experiments, analyzed the data, and wrote the original draft of the manuscript. Sukanya Maity was responsible for the synthesis of the materials. Mustafa Mahmoud Aboulsaad and Tomas Edvinsson carried out the *in situ* Raman spectroscopy experiments, and Mustafa Mahmoud Aboulsaad wrote the *in situ* Raman spectroscopy section of the manuscript. Johanna Rosen reviewed the manuscript, Per O.Å. Persson contributed to experimental planning, manuscript writing, and critical revision. All authors reviewed the manuscript, contributed to its preparation, and approved the final version.

**FUNDING SOURCES**

The Wallenberg Initiative for Sustainable Materials (WISE-AP01-D41),

The Swedish Research Council (2021-04499, 2021-00171),

The Swedish Energy Agency for project grants (P2020-90149),

The Swedish Foundation for Strategic Research (RIF21-0026),

The Swedish Government Strategic Research Area (Faculty Grant SFO-Mat-LiU No. 2009-00971).

MMA and TE acknowledge funding from the Swedish Research council (2023-05244).

The Swedish Energy Agency (P2020-90215).

The European Union (ERC, MULT12D, 101087713)

The KAW foundation (KAW 2023,0250, KAW 2020,0033)

**NOTES**

The authors declare no competing financial interest

**ACKNOWLEDGMENT**

The authors thank the Wallenberg Initiative for Sustainable Materials (WISE) for a project grant (WISE-AP01-D41), The Swedish Research Council (VR, 2023-05244 and The Swedish Energy Agency for project grants (2021-04499, P2020-90149 and P2020-90215), respectively. The Knut and Alice Wallenberg (KAW) Foundation is acknowledged for support of the Linköping Electron Microscopy Laboratory, for a Scholar grant (KAW 2023, 0250) and project funding (KAW 2020,0033). The European Union is acknowledged for a Consolidator grant (ERC, MULTI2D, 101087713). VR and the Swedish Foundation for Strategic Research (SSF) are further acknowledged for access to ARTEMI, the Swedish National Infrastructure in Advanced Electron Microscopy (2021-00171 and RIF21-0026). The authors also acknowledge the Swedish

Government Strategic Research Area in Materials Science on Advanced Functional Materials at Linköping University (Faculty Grant SFO-Mat-LiU No. 2009-00971).

**REFERENCES**

(1) Shanmugam, V.; Mensah, R. A.; *et al.* A Review of the Synthesis, Properties, and Applications of 2D Materials. *Part. Part. Syst. Charact.* **2022**, *39* (6), 2200031.

(2) Ge, M.; Cao, C.; *et al.* A Review of One-Dimensional $TiO_2$ Nanostructured Materials for Environmental and Energy Application. *J. Mater. Chem. A* **2016**, *4*, 6772–6801.

(3) Katan, C.; Mercier, N.; Even, J. Quantum and Dielectric Confinement Effects in Lower-Dimensional Hybrid Perovskite Semiconductors. *Chem. Rev.* **2019**, *119* (5), 3140–3192.

(4) Wang, J.; Jiang, C.; Li, W.; Xiao, X. Anisotropic Low-Dimensional Materials for Polarization-Sensitive Photodetectors: From Materials to Devices. *Adv. Opt. Mater.* **2022**, *10* (6), 2102436.

(5) Yu, Y.; Zhuo, M.-P.; Chen, S.; He, G.-P.; Tao, Y.-C.; Wang, X.-D.; Liao, L.-S. Molecular- and Structural-Level Organic Heterostructures for Multicolour Photon Transportation. *J. Phys. Chem. Lett.* **2020**, *11* (18), 7517–7524.

(6) Das, C. M.; Kang, L.; Ouyang, Q.; Yong, K.-T. Advanced Low-Dimensional Carbon Materials for Flexible Devices. *InfoMat* **2020**, *2* (4), 698–714.

(7) Cao, M.-S.; Wang, X.-X.; Zhang, M.; Shu, J.-C.; Cao, W.-Q.; Yang, H.-J.; Fang, X.-Y.; Yuan, J. Electromagnetic Response and Energy Conversion for Functions and Devices in Low-Dimensional Materials. *Adv. Funct. Mater.* **2019**, *29* (25), 1807398.

(8) Ogletree, D. F. Revealing Optical Properties of Reduced-Dimensionality Materials at Relevant Length Scales. *Adv. Mater.* **2015**, *27* (38), 5696–5701.

(9) Tong, X.; Zhan, X.; Rawach, D.; Chen, Z.; Zhang, G.; Sun, S. Low-Dimensional Catalysts for Oxygen Reduction Reaction. *Prog. Nat. Sci.: Mater. Int.* **2020**, *30* (6), 787–795.

(10) Xie, B.; Zhu, Y.; Marwat, M. A.; Zhang, S.; Zhang, L.; Zhang, H. Tailoring the Energy Storage Performance of Polymer Nanocomposites with Aspect Ratio Optimized 1D Nanofillers. *J. Mater. Chem. A* **2018**, *6* (41), 20356–20364.

(11) Mendoza-Sánchez, B.; *et al.* Synthesis of Two-Dimensional Materials for Capacitive Energy Storage. *Adv. Mater.* **2016**, *28* (29), 6104–6135.

(12) Maity, S.; Qin, L.; Petruhins, A.; Barsoum, M. W.; Rosén, J. One-Dimensional Lepidocrocite Titania Mesoparticles Integrated with Activated Carbon for High-Performance Supercapacitor Applications. *J. Energy Storage* **2024**, *101*, 113895.

(13) Weng, B.; Liu, S.; Tang, Z.-R.; Xu, Y.-J. One-Dimensional Nanostructure Based Materials for Versatile Photocatalytic Applications. *RSC Adv.* **2014**, *4* (25), 12685–12700.

(14) Zang, L.; Che, Y.; Moore, J. S. Construction and Optoelectronic Properties of Organic One-Dimensional Nanostructures. *Acc. Chem. Res.* **2008**, *41* (12), 1596–1608.

(15) Wang, L.; Badr, H. O.; Yang, Y.; Cope, J. H.; Ma, E.; Ouyang, J.; Yuan, L.; Li, Z.; Liu, Z.; Barsoum, M. W.; Shi, W. Unique Hierarchical Structures of One-Dimensional Lepidocrocite Titanate with Cation-Exchangeable Sites for Extraordinary Selective Actinide Capture for Water Purification. *Chem. Eng. J.* **2023**, *474*, 145635.

(16) Shin, J.; Kang, N.; Kim, B.; Hong, H.; Yu, L.; Kim, J.; Kang, H.; Kim, J. S. One-Dimensional Nanomaterials for Cancer Therapy and Diagnosis. *Chem. Soc. Rev.* **2023**, *52* (13), 4488–4514.

(17) Sorooshian, S. The Sustainable Development Goals of the United Nations: A Comparative Midterm Research Review. *J. Cleaner Prod.* **2024**, 142272.

(18) Maity, S.; Ibrahim, M.; Badr, H.; Reji, T.; Hassig, M. Q.; Zhang, T.; Li, C.; Rosén, J.; Barsoum, M. W. Titanium Oxysulfate-Derived 1D Lepidocrocite Titanate Nanostructures. *Adv. Mater. Interfaces* **2024**, *12* (10), 2400866.

(19) Tseng, E. N.; Iythichanda Risha, A.; Alnoor, H.; Zheng, W.; Huang, W. Z.; Rosén, J.; Persson, P. O. Å. Exploring 1D and 2D Atomically Thin Lepidocrocite TiO₂. *Submitted*.

(20) Thakur, N.; Thakur, N.; Kumar, A.; *et al.* A Critical Review on the Recent Trends of Photocatalytic, Antibacterial, Antioxidant and Nanohybrid Applications of Anatase and Rutile TiO₂ Nanoparticles. *Sci. Total Environ.* **2024**, *914*, 169815.

(21) Zhu, S.-C.; Xie, S.-H.; Liu, Z.-P. Nature of Rutile Nuclei in Anatase-to-Rutile Phase Transition. *J. Am. Chem. Soc.* **2015**, *137* (35), 11532–11539.

(22) Che, X.; Li, L.; Zheng, J.; Li, G.; Shi, Q. Heat Capacity and Thermodynamic Functions of Brookite TiO₂. *J. Chem. Thermodyn.* **2016**, *93*, 45–51.

(23) Gao, T.; Norby, P.; Okamoto, H.; Fjellvåg, H. Syntheses, Structures, and Magnetic Properties of Nickel-Doped Lepidocrocite Titanates. *Inorg. Chem.* **2009**, *48* (19), 9409–9418.

(24) Gao, T.; Fjellvåg, H.; Norby, P. Crystal Structures of Titanate Nanotubes: A Raman Scattering Study. *Inorg. Chem.* **2009**, *48* (4), 1423–1432.

(25) Tian, F.; Zhang, Y.; Zhang, J.; Pan, C. Raman Spectroscopy: A New Approach to Measure the Percentage of Anatase $TiO_2$ Exposed (001) Facets. *J. Phys. Chem. C* **2012**, *116* (13), 7515–7519.

(26)  Hu, W.; Li, L.; Li, G.; Liu, Y.; Withers, R. L. Atomic-Scale Control of $TiO_6$ Octahedra through Solution Chemistry towards Giant Dielectric Response. *Sci. Rep.* **2014**, *4*, 6582.

(27) Schwenk, G. R.; Walter, A. D.; Barsoum, M. W. Solvent-Driven Self-Assembly of One-Dimensional Lepidocrocite Titanium-Oxide-Based Nanofilaments. *Nano Lett.* **2024**, *24* (25), 7584–7592.

(28) Bak, T.; Nowotny, J.; Nowotny, M. K. Defect Disorder of Titanium Dioxide. *J. Phys. Chem. B* **2006**, *110* (43), 21560–21567.

(29) Su, J.; Zou, X.; Chen, J.-S. Self-Modification of Titanium Dioxide Materials by $Ti^{3+}$ and/or Oxygen Vacancies: New Insights into Defect Chemistry of Metal Oxides. *RSC Adv.* **2014**, *4*, 13979–13988.

(30) Khatim, O.; Amamra, M.; Chhor, K.; Bell, A. M. T.; Novikov, D.; Vrel, D.; Kanaev, A. Amorphous–Anatase Phase Transition in Single Immobilized $TiO_2$ Nanoparticles. *Chem. Phys. Lett.* **2013**, *558*, 53–56.

(31) Baranovskyi, D.; Kolesnichenko, V.; Tyschenko, N.; Lobunets, T.; Shyrokov, O.; Baranovska, O.; Ragulya, A. Crystallization Kinetics of Nano-Sized Amorphous $TiO_2$ during Transformation to Anatase under Non-Isothermal Conditions. *Next Mater.* **2025**, *9*, 101106.

**Table of Contents (TOC)**

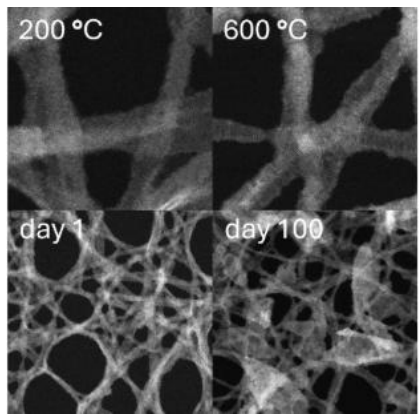

The thermal and environmental stability of one-dimensional (1D) lepidocrocite titanium dioxide (TiO$_2$) filaments is systematically explored. These nanostructures retain their structural integrity up to 300 °C in vacuum while they undergo amorphization followed by sintering and finally transformation into the anatase structure under extended heating. Extended exposure (> 100 days) to ambient conditions while in aqueous solution also result in phase transformation of the filaments into anatase TiO$_2$, although refrigerated conditions significantly reduce the transformation. Accordingly, this investigation thus explores the intrinsic boundary conditions for this material in colloidal and high temperature applications.